\documentclass[prd,onecolumn,notitlepage,showpacs,preprintnumbers,amsmath,amssymb,nofootinbib,APS,10pt,superscriptaddress]{revtex4-1}

\usepackage{dcolumn}
\usepackage{bm}
\usepackage{xcolor}
\usepackage[utf8]{inputenc}
\usepackage[spanish,english]{babel}
\usepackage{amsmath,amssymb,amsfonts,latexsym,cancel}
\usepackage{graphicx}
\usepackage{color}
\usepackage{soul}
\usepackage{ulem}
\usepackage{hyperref}
\usepackage{amsmath}
\usepackage{slashed}
\usepackage{braket}
\usepackage{amssymb}
\usepackage{amsmath}

\usepackage{graphicx}
\graphicspath{ {./images/} }
\allowdisplaybreaks

\newcommand{\be}{\begin{equation}}
\newcommand{\ee}{\end{equation}}
\newcommand{\bea}{\begin{eqnarray}}
\newcommand{\eea}{\end{eqnarray}}

\begin{document}

\title{{\bf  
Adiabatic regularization and preferred vacuum state for the $\lambda \phi^4$ field theory in cosmological spacetimes}} 

\author{Antonio Ferreiro}\email{antonio.ferreiro@dcu.ie}
\affiliation{Centre for Astrophysics and Relativity, School of Mathematical Sciences,
Dublin City University, Glasnevin, Dublin 9, Rep. of Ireland.}

\author{Silvia Pla}\email{silvia.pla@uv.es}
\affiliation{Departamento de Fisica Teorica and IFIC, Centro Mixto Universidad de Valencia-CSIC. Facultad de Fisica, Universidad de Valencia, Burjassot-46100, Valencia, Spain.}

\begin{abstract}
 We extend the method of adiabatic regularization by introducing an arbitrary parameter $\mu$ for an scalar field with quartic self-coupling in a Friedmann-Lemaître-Robertson-Walker (FLRW) spacetime at one-loop order. The subtraction terms constructed from this extended version allow us to define a preferred vacuum state at a fixed time $\eta=\eta_0$ for this theory. We compute this vacuum state for two commonly used background fields in cosmology. We also give a possible prescription for an adequate value for $\mu$. \newline

\end{abstract}
\date{\today}
\maketitle

\section{Introduction}

The construction of a theory of quantum fields propagating in  curved spacetimes has been a very fruitful endeavour both in its theoretical-mathematical formulation and in its astrophysical and cosmological applications \cite{Birrel, Fulling, Wald, Navarro, Parker, Hu, Shapiro}. One of the main lessons we have learned so far is that the traditional concepts in flat spacetime of a preferred vacuum state and normal ordering are not to be taken for granted anymore. In this context, the construction of the expectation value of the stress-energy tensor is highly non trivial. 
Accordingly, it is necessary to construct this magnitude consistently, not only to determine the local energy, momentum, and stress properties of the quantized field but also because it 
plays a crucial role in the semiclassical Einstein equations, which describe the backreaction of the quantum field on the spacetime geometry. In the case of a cosmological spacetime, we can perform a Fourier transformation on the fields and carry out a mode sum expansion. The relative simplicity of the equations and the construction of the stress-energy tensor in this description has been not only fruitful for cosmological applications but also for conceptual understanding of quantum properties in non-flat spacetimes  \cite{Parkerthesis, Birrel, Fulling, Parker, Hu}. \\

A well-known difficulty in the construction of the vev of the stress-energy tensor is the presence of ultraviolet (UV) divergences, where normal-ordering is not applicable anymore. Many methods, known as regularization and renormalization, have been developed to overcome these infinities and produce finite physical results \cite{Birrel, Fulling, Wald, Parker,DeWitt75,Christensen,BunchParker,Wald77}. In the case of a Friedman-Lemaître-Robertson-Walker (FLRW) spacetime, adiabatic regularization constructs the subtraction terms in such a way that they include all possible divergences of the stress-energy tensor while maintaining the mode sum description and therefore 
preserving its computational  
efficiency \cite{Parkerthesis, Adiabatic1}. Even though the spacetime geometry is fixed to be FLRW, it is important to stress out that this regularization method is equivalent to more general methods such as the DeWitt-Schwinger 
asymptotic expansion when restricting to an homogeneous and isotropic geometry \cite{DN15, BDNN21} (see also \cite{Birrell78}). Adiabatic regularization has been successfully extended to spin-$\tfrac{1}{2}$ fields \cite{Adiabatic2} and to 
classical scalar and electromagnetic background fields \cite{DFNT,Adiabatic3}.\\

In general, there exists an arbitrariness in the construction of a regularization and renormalization program. For example, in dimensional regularization, this arbitrariness is usually encoded in a mass parameter $\mu$ \cite{Hooft1,Hooft2}. Of course, there is no harm in this ambiguity since the difference between two different values for $\mu$ can always be reabsorbed into the renormalized coupling constants. A change in the renormalization point $\mu$  corresponds to a change in the renormalized coupling constants. In flat spacetime, this change or running of the coupling constants has been proven to be very fruitful, for example, to analyze the behavior of gauge theories in a high energy limit or to construct effective potentials for the quantized fields
 \cite{coleman}.\\

In curved spacetime, there also exists an arbitrariness when constructing the subtraction terms. For example, for a free massive scalar field in a four-dimensional curved spacetime, the possible difference between two stress-energy tensors with different renormalization schemes yields \cite{Wald}
\bea
\langle T_{ab}\rangle_{\textrm{ren}}-\langle \tilde{T}_{ab}\rangle_{\textrm{ren}}=a g_{ab}+b G_{ab}+c H_{ab}^{(1)}+d H_{ab}^{(2)},
\eea
where $G_{ab}$ is the Einstein tensor and $H_{ab}^{(1)}$ and $H_{ab}^{(2)}$ are tensors constructed from higher-order curvature terms. In the case of adiabatic regularization, it has been shown that one possibility is to encode  this arbitrariness into a mass scale $\mu$, similar to the dimensional regularization analogue, by upgrading the leading order of the 
adiabatic expansion from $\omega^{(0)}=\sqrt{k^2+a^2m^2}$ to $\tilde{\omega}^{(0)}=\sqrt{k^2+a^2\mu^2}$, where $m$ is the physical mass of the field \cite{FN1}. The consequences of this extended adiabatic regularization method have been recently studied in \cite{Sola1} and \cite{Sola2} in the context of the Running Vacuum Model and the cosmological constant problem. In this paper, we generalize this extended adiabatic regularization method  for the case of an interacting $\lambda \phi^4$ theory in an expanding universe. Note that the standard subtraction terms constructed via adiabatic regularization \cite{Adiabatic1} 
can be regarded as a particular scheme, i.e.,  $\mu^2=m^2$.\\

Another difficulty in Quantum Field Theory in curved spacetimes is the absence of a preferred vacuum state, even in the case of isotropic and homogeneous spacetimes. Indeed, any state that is Hadamard or adiabatic for cosmological spacetimes is equally suitable for the definition of a vacuum state \cite{Wald, Fulling}. 
Therefore, to select an adequate vacuum state, additional requirements need to be 
considered,  e.g., taking advantage of the emergent
conformal symmetry near the big bang in a radiation-dominated universe \cite{Beltran}. 
For general FLRW spacetimes, a recent method to select an adequate vacuum state was proposed in \cite{Agullo15} for the case of a free linear scalar field. This vacuum state verified the necessary restrictions by requiring that the mode expansion of the regularized stress-energy tensor vanish mode-by-mode at a given time $\eta=\eta_0$. It was also proven that this condition is sufficient to uniquely determine the vacuum state for several relevant backgrounds in cosmology.\\ 

An interesting question is whether this requirement can be imposed into an interacting theory, e.g., a scalar field with an $\lambda \phi^4$ potential. This question is of special importance since it is well-known that scalar fields with a potential play an essential role in the dynamics of the early stages of the Universe \cite{reheating}. Furthermore, in order to quantify the possible quantum  production of particles due to the scalar background field \cite{preheating}, we need to be able to both construct a regularize stress-energy tensor and select a preferred vacuum state.  As we will show in this work, the above-mentioned requirement for the regularized stress-energy tensor via standard adiabatic regularization fails to produce well-defined modes for physical models of inflation and reheating.\\

Since this vacuum state is constructed from the subtraction terms of adiabatic regularization, different schemes can result in different vacuum states. In the case of the extended version of adiabatic regularization \cite{FN1}, each possible value for $\mu$ will produce a different vacuum state. We will show that there is always the possibility to choose a specific value of $\mu$ such that there exist modes of the vacuum state that have vanishing stress-energy tensor mode-by-mode at a given time $\eta=\eta_0$. This is a generalization of the results in \cite{Agullo15}, since it is a particular case for $\mu=m$ and $\lambda=0$. \\

The paper is organized as follows. In Section \ref{sec:lambdaphi4} we  introduce the model that we  use through the rest of the paper. We also give the formal expressions  of some relevant quantities, such as the energy density or the pressure. In Section \ref{sec:adiabatic} we  introduce the $\mu$-extended version of the adiabatic regularization method when including scalar background fields. We  also give the renormalized vacuum expectation values of the main observables of the theory. In Section \ref{sec:instantaneousvacuum} we  introduce an upgraded version of the instantaneous vacuum proposed in \cite{Agullo15} including scalar background fields and for an arbitrary $\mu$. For the $\lambda \phi^4$ potential in a FLRW universe, we deeply study the region of validity of our instantaneous vacuum and its dependence on the $\mu$ parameter. Finally, in Section \ref{sec:daughterfield} we will re-analyze the validity of the instantaneous vacuum as a function of $\mu$ for the potential $\frac{1}{2}g^2 \phi^2 X^2$, that is, for a daughter field $X$ coupled with the usual inflaton field $\phi$. In Section \ref{sec:conclusions} we give some concluding remarks.
\section{Scalar field in a FLRW spacetime}
\label{sec:lambdaphi4}

Consider the action of a scalar field $\phi$, non-minimally coupled to the curvature 
\be S_{\phi}[\phi,g_{\mu\nu}] = \frac12 \int d^4x \sqrt{-g} \left(  g^{\mu \nu} \partial_{\mu} \phi \partial_{\nu} \phi -( \xi R+m^2) \phi^2-\frac{\lambda}{2} \phi^4   \right) \ . \label{eq:scalar-action} \ee
Here $m^2$, $\lambda$, $\xi$ are the bare mass, quartic coupling and scalar curvature coupling respectively. The associated Klein-Gordon equation for the scalar field is
\be
(\Box+\xi R+m^2+\lambda \phi^2)\phi=0, \label{KGphi}
\ee
and the corresponding stress-energy tensor reads
\bea
T_{ab}=\nabla_a\phi\nabla_b\phi-\frac12 g_{ab}\nabla^c\phi\nabla_c\phi+g_{ab}\left(\frac{m^2}{2} \phi^2+\frac{\lambda}{4} \phi^4\right)-\xi\left(R_{ab}-\frac12 R g_{ab}\right)\phi^2+\xi \left(g_{ab}\nabla^c\nabla_c\phi^2-\nabla_a\nabla_b\phi^2\right).\label{Tab}
\eea
It is useful to break the field into its mean field $\bar{\phi}=\langle \phi \rangle$ and the fluctuation field $  \delta \phi $ as $ \phi  \equiv \bar{\phi} +  \delta \phi \label{eq:FieldFluct}$.
Following \cite{Ringwald1, Anderson1} and truncating at one-loop order, the equations of motion become
\bea
\left(\Box+\xi R+m^2+\lambda \bar{\phi}^2+3\lambda \langle \delta \phi^2\rangle\right)\bar \phi=0, ~~~~~~~ \left(\Box+\xi R+m^2+3\lambda \bar{\phi}^2\right)\delta \phi=0.\label{KGfinalphi}
\eea
where $\langle\delta \phi^2 \rangle \equiv \bra{0}\delta \phi^2 \ket{0}$ is the vacuum expectation value of the fluctuating field. The same equations can be obtained by making use of the $1/N$ approximation (see \cite{Hu} for a detailed explanation). 
The semiclassical Einstein equation at one-loop order is
\bea 
(8\pi G)^{-1}G_{ab}+\Lambda g_{ab}+\alpha ^{(1)}H_{ab}+\beta ^{(2)}H_{ab}=8\pi G\left(\bar{T}_{ab}(\bar{\phi})+\langle T_{ab}\rangle\right).
\eea
$\bar{T}_{ab}(\bar{\phi})$ is the stress-energy tensor \eqref{Tab} for $\bar{\phi}$ and $\langle T_{ab}\rangle\equiv \bra{0}T_{ab} \ket{0} $ is the vacuum expectation value of the stress-energy tensor of a free field with effective square mass $Q:= m^2+3\lambda \bar{\phi}^2$. Both $ \langle \delta \phi^2\rangle$ and $\langle T_{ab}\rangle$ are divergent and need to be regularized and renormalized by adding the corresponding counter-terms. The $\lambda \phi^4$ theory can be renormalized in general curved spacetime \cite{Bunch1, Bunch2}. The renormalization involves shifting the bare coupling constants of \eqref{eq:scalar-action} to their renormalized, finite analogs. For example, in dimensional regularization, the relation between the bare couplings and the renormalized ones are
\bea
&&m^2\equiv m_R^2-\frac{3\lambda_R}{8\pi^2 (n-4)}m_R^2\\
&&\xi-\frac16 \equiv \xi_R-\frac16 -\frac{3\lambda_R}{8\pi^2 (n-4)}(\xi_R-\frac16)\\
&&\lambda\equiv \lambda_R-\frac{9\lambda^3_R}{8\pi^2(n-4)}, 
\eea
while the renormalized semiclassical equations are \cite{Anderson2, Hu}
\bea
\left(\Box+\xi_{\rm R} R+m_{\rm R}^2+\lambda_{\rm R} \bar{\phi}^2+\lambda_{\rm R} \langle \delta \phi^2\rangle_{\rm ren}\right)\bar \phi=0, ~~~~~~~ \left(\Box+\xi_{\rm R} R+m_{\rm R}^2+3\lambda_{\rm R} \bar{\phi}^2\right)\delta \phi=0.\label{KGfinalphi}
\eea
\bea \label{semiren}
(8\pi G_{\rm R})^{-1}G_{ab}+\Lambda_{\rm R} g_{ab}+\alpha_{\rm R} ^{(1)}H_{ab}+\beta_{\rm R} ^{(2)}H_{ab}=\left(\bar{T}_{ab}(\bar{\phi})+\langle T_{ab}\rangle_{\rm ren}\right).
\eea
 From now on, we drop the $_{\rm R}$ index for simplicity, i.e., all the couplings constants appearing into the computations are the finite renormalized couplings. In the next section, we show how to construct the finite magnitudes $\langle \delta \phi^2\rangle_{\rm ren}$ and $\langle T_{ab}\rangle_{\rm ren}$. \\

In the case of a flat FLRW metric $ds^2=a(\eta)^2\left(d\eta^2-d\vec{x}^2\right)$, where $\eta$ is the conformal time,  we can assume $\bar \phi=\bar \phi(\eta)$ and express the 
quantum fluctuations as 
\bea
\delta\phi(\vec{x},\eta)=\frac{1}{\left(2\pi\right)^3}\int d^3 k \left[A_{\vec{k}} h_{\vec{k}}(\eta)e^{i \vec{k}\vec{x}}+A_{\vec{k}}^{\dagger}h_{\vec{k}}^{*}(\eta) e^{-i \vec{k}\vec{x}}\right].\label{deltaphimod}
\eea
By choosing the normalization conditions $h_{\vec{k}}h^{*\prime}_{\vec{k}}-h_{\vec{k}}'h^{*}_{\vec{k}}=ia^{-2}$ and $h_{\vec{k}}h^{\prime}_{-\vec{k}}-h_{\vec{k}}'h_{-\vec{k}}=0$, the operators $A_{\vec{k}}$ and $A_{\vec{k}}^{\dagger}$ can be interpreted as the usual creation and annihilation operators.
At this point, we can define the vacuum state $\ket{0}$ as the state annihilated by the operator $A_{\vec{k}}$. The modes $h_{\vec{k}}$ follow the equation of motion 
\bea
h_{\vec k}''+2\frac{a'}{a}h'_{\vec k} +\left(k^2+a^2 Q + 6\xi \frac{a''}{a}\right)h_{\vec k}=0,\label{eom}
\eea
where here again $Q\equiv m^2+3\lambda \bar{\phi}^2$. We note that the expected invariance under rotations requires that the modes $h_{\vec k} $ depend only on $k=|\vec k|$. Therefore, from now on we drop the vector $\vec{k}$ and write $k=|\vec{k}|$. In an isotropic and homogeneous spacetime, the vacuum expectation value of the stress-energy tensor can be decomposed in terms of 
\bea
\langle T_{ab}\rangle=-g_{ab}\langle p\rangle+(\langle p\rangle+\langle \rho\rangle )u_a u_b \label{Trhop}
\eea
where $u^a$ is the unit vector normal to the homogeneous and isotropic hypersurface. In terms of mode function of \eqref{deltaphimod}, the components of \eqref{Trhop} become

\bea
\langle \rho \rangle\equiv\frac{1}{(2\pi)^3}\int d^3k \langle \rho_k\rangle=&&\frac{1}{(2\pi)^3}\int \frac{d^3k}{2a^2} \left(|h'_{k}|^2+\left(k^2+Qa^2\right)|h_{k}|^2+6\xi\left(\frac{a'^2}{a^2}|h_{k}|^2+\frac{a'}{a}(h_{k}h'^{*}_{k}+h^{*}_{k}h'_{k})\right)\right)\label{rhok}\\
\langle p \rangle\equiv\frac{1}{(2\pi)^3}\int d^3k \langle p_k\rangle=&&\frac{1}{(2\pi)^3}\int \frac{d^3k}{2a^2} \Bigg(|h'_{k}|^2-\left(\frac{k^2}{3}+Qa^2\right)|h_{k}|^2-2\xi\left(\left(2-12\xi\right)\frac{a''}{a}-\frac{a'^2}{a^2}\right)|h_{k}|^2\nonumber \\&&+2\xi\left(\frac{a'}{a}(h_{k}h'^{*}_{k}+h^{*}_{k}h'_{k})-2|h'_{k}|^2+\left(2k^2+2Qa^2\right)|h_{k}|^2\right)\Bigg).\label{pk}
\eea
As we have already stressed out, both of these quantities diverge. In the next section, we will use adiabatic regularization to construct finite energy density \eqref{rhok} and pressure \eqref{pk}.

\section{Adiabatic Regularization} \label{sec:adiabatic}
We briefly introduce how to perform adiabatic regularization (for a more extended description see \cite{Birrel, Parker}). First, we introduce the WKB ansatz 
\bea
h_{k}\sim \frac{1}{a\sqrt{2W_{k}}}e^{-i\int^\eta W_{ k} d\eta'}\label{eomh}
\eea
into the equations of motion \eqref{eom} which results in (we drop the sub-index $k$ for convenience)
\bea
W^{1/2}\frac{d^2}{d\eta^2}W^{-1/2}+\left(k^2+a^2Q\right)+\left(6\xi-1\right)\frac{a''}{a}=W^2.\label{equW}
\eea
We want to obtain an (adiabatic) expansion of $W=\omega^{(0)}+\omega^{(1)}+\omega^{(2)}+...$ where each term $\omega^{(n)}$ has a fixed adiabatic order $n$. To this end,  it becomes necessary to give a prescription of the adiabatic order for each time-dependent parameter. In the case of the scale factor $a(\eta)$, which is of adiabatic order zero, each derivative gives rise to an extra adiabatic order \cite{Parker}. The case of $Q$ is more subtle. A first possibility is to assume $Q$ of adiabatic order zero. This is the standard procedure for the free field case when $Q=m^2$ is a constant \cite{Parker}. However, if $Q$ is not constant, e.g. for a polynomial type potential $V \propto \phi^n$ with $n>2$ with $Q\propto \bar{\phi}^{n-2}$, then the adiabatic order zero assignation is no longer valid. In \cite{Anderson1} for the special case of $n=4$ it was found that the consistent adiabatic order for $Q$ is order two (see also \cite{DFNT}). Therefore, it is reasonable to choose $Q$ of adiabatic order two. \\

In this case, we need to introduce an additional parameter $\mu^2$ to avoid infrared divergences. This is equivalent to the infrared divergences appearing in the massless case for free fields \cite{Parker, FN1}. We follow the approach used in \cite{FN1}. Equation \eqref{equW} is modified as follows
\bea
W^{1/2}\frac{d^2}{d\eta^2}W^{-1/2}+\left(k^2+a^2\mu^2\right)+\left(a^2Q-a^2\mu^2\right)+\left(6\xi-1\right)\frac{a''}{a}=W^2. \label{equW2}
\eea
Here $\left(k^2+a^2\mu^2\right)=:\omega^2$ is assumed of adiabatic order zero, while $\left(a^2Q-a^2\mu^2\right)+\left(6\xi-1\right)\frac{a''}{a}:=\sigma$ is assumed of adiabatic order two.\footnote{Note that an alternative is to consider only the $\mu^2$ in the first parenthesis of \eqref{equW2} and then take, at the end of the calculation, $\mu^2\to 0$. We do not follow this approach here for practical purposes as we will see in the next section} We can solve \eqref{equW} iteratively, which results in 
\bea
&&\omega^{(0)}=\omega,~~~~~\omega^{(2)}=\frac12 \omega^{-1/2}\frac{d^2}{d\eta^2}\omega^{-1/2}+\frac12 \omega^{-1}\sigma\\
&&\omega^{(4)}=\frac14 \omega^{(2)}\omega^{-3/2}\frac{d^2}{d\eta^2}\omega^{-1/2}-\frac12 \omega^{-1}\left(\omega^{(2)}\right)^2-\frac14 \omega^{-1/2}\frac{d^2}{d\eta^2}\left(\omega^{-3/2}\omega^{(2)}\right).
\eea
We only need to compute up to adiabatic order four since the subtractions for the stress-energy tensor are required only up to 
this order. For the renormalized two point function we obtain
\bea \label{eq:tpf}
\langle  \delta \phi^2 \rangle_{\rm ren}=\frac{1}{(2\pi)^3}\int d^3k |h_{{k}}|^2-\Phi^{(0)}-\Phi^{(2)}
\eea
where 
\bea 
\Phi_{\rm}^{(0)}=\frac{1}{(2\pi)^3}\int d^3k \frac{1}{2a^2\omega}\,,~~~\Phi_{\rm}^{(2)}=\frac{-1}{(2\pi)^3}\int d^3k\frac{\omega^{(2)}}{2a^2\omega^2}.\label{phiadb}
\eea
The two-point function needs only to be subtracted up to adiabatic order two since the fourth adiabatic order is already finite.
In the case of the stress-energy tensor we need to subtract up to adiabatic order four
\bea
\langle T_{ab}\rangle_{\rm ren}=\langle T_{ab}\rangle-\mathcal{T}_{ab}^{(0)}-\mathcal{T}_{ab}^{(2)}-\mathcal{T}_{ab}^{(4)},
\eea
and in terms of the individual components \eqref{Trhop} 
\be 
\langle \rho \rangle_{\rm ren}=\frac{1}{(2\pi)^3}\int d^3k \Big[\langle \rho_k\rangle-C_{\rho}(\mu,k,\eta)\Big] \, , \quad \langle p \rangle_{\rm ren}=\frac{1}{(2\pi)^3}\int d^3k \Big[\langle p_k \rangle-C_{p}(\mu,k,\eta)\Big].\label{rhop}
\ee
The subtractions for each component $C_{\rho}$ and $C_{p}$ can be found in Appendix \ref{app:subtractions}.



\subsection{$\mu$-invariance of the Renormalized Semiclassical Equations}

It is important to check whether the introduction of $\mu$ does not generate additional counter-terms into the renormalized semiclassical equations \eqref{KGfinalphi} and \eqref{semiren}. There are different ways of proving this. One option is to compute the difference between two stress-energy tensors, obtained via adiabatic regularization with two different prescriptions $\mu_1$ and $\mu_2$. This was done in \cite{FN1} for the case of a scalar field with an electromagnetic field. Here we follow a different approach. We take the stress-energy tensor renormalized with a general prescription $\mu$ and perform the derivative with respect to this parameter. It is easy to check that, at one-loop approximation, this magnitude gives a finite quantity
\bea
\mu\frac{d}{d\mu}\langle T_{ab}\rangle_{\rm ren}=a g_{ab}+b\left(\xi-\frac16\right)G_{ab}+c\left(\xi-\frac16\right)^2 {}^{(1)}H_{ab}+S_{ab},\label{dmuT}
\eea
where $a$, $b$ and $c$ are functions of $\mu^2$, and $S_{ab}$ is defined as
\bea
S_{ab}=\frac{1}{16\pi^2}\left(-\frac12Q^2 g_{ab}+\mu^2 Q g_{ab}+ \left(6\xi-1\right)QG_{ab}-\left(1-6\xi\right)\left(g_{ab}\nabla^c \nabla_c Q-\nabla_a \nabla_b Q\right)\right).
\eea
For our particular case $Q=m^2+3\lambda \bar{\phi}^2$, all  terms appearing in \eqref{dmuT} are already present in the renormalized semiclassical equations  \eqref{KGfinalphi} and \eqref{semiren}. Therefore, any possible difference between the reparametrization of an initial prescription of the subtraction terms $\mu=\mu_1$ into another $\mu=\mu_2$ can always be reabsorbed by reparametrizing the coupling constants of the renormalized semiclassical equations  \eqref{KGfinalphi} and \eqref{semiren}.\\

By requiring $\mu$-independence of the semiclassical equations  \eqref{KGfinalphi} and \eqref{semiren}, the renormalization of the coupling constant generates an effective dependence on the $\mu$ scale. In theories where we can assign a physical interpretation to $\mu$ as an energy scale, one can study the behavior of the couplings and, therefore, of the theory at some extreme high or low energy regime. This lies at the heart of the effective field theory approach or the asymptotic free theories \cite{QFT1, QFT2,QFT3}. Some attempts have been carried out to study this possible scale dependence in general curved spacetime \cite{Nelson-Panangaden, Parker-Toms, Shapiro} and for the expansion of the Universe, \cite{Shapiro2, Babic}. Nevertheless, the possible interpretations in this context are still not completely understood. Here, we will carry on a different approach. We will fix the arbitrary parameter $\mu=\mu_{*}$ in order to obtain the semiclassical equations  \eqref{KGfinalphi} and \eqref{semiren} with fixed coupling constants, which would be determined by observations. Therefore, once we have fixed the corresponding value of $\mu_*$, the renormalized couplings constants will not change.  \\

In the next section, we obtain a vacuum state that produces vanishing renormalized vacuum expectation values of the pressure and the density  \eqref{rhop} mode by mode at a given time $\eta_0$. Since the subtraction terms $C_{\rho}$ and $C_{p}$ will be prescribed with $\mu=\mu_{*}$, the terms $\langle \rho_k \rangle_{\textrm{ren}}$ and $\langle p_k \rangle_{\textrm{ren}}$ will also inherit a dependence of $\mu_{*}$. It is important to note that once this parameter has been fixed, it can not be arbitrarily changed in future calculations without changing the coupling constants accordingly.\\

We use this dependence with $\mu_{*}$ to construct a well-defined vacuum state. We show that not any arbitrary value of $\mu_{*}$ allows us to construct a well-defined vacuum state. Nonetheless, it is always possible to construct it for interesting physical scenarios.

\section{Instantaneous Vacuum for the minimal coupled scalar field} \label{sec:instantaneousvacuum}

In order to construct the vacuum state, we follow the same approach as in \cite{Agullo15} for the case of a minimally coupled scalar field. It can be done as follows. To choose a vacuum at $\eta=\eta_0$ implies to choose a particular set of initial conditions $\{h_{ k}(\eta_0), h'_{ k}(\eta_0)\}$. In this context, they can be conveniently parametrized as 
\bea \label{IV-1}
h_{ k}(\eta_0)=\frac{1}{a(\eta_0)\sqrt{2 W_{k}(\eta_0)}}\, , \qquad h'_{ k}(\eta_0)=\Big(-i W_{ k}(\eta_0)+\frac{V_{ k}(\eta_0)}{2}-\frac{a'(\eta_0)}{a(\eta_0)}\Big)h_{ k}(\eta_0).\label{modeansatz}
\eea
To ensure that the solutions are normalized, 
  both $W_{k}$  and $V_{ k}$ have to be real, and $W_{k}$ has to be also positive for all $ k$. 
  As we have already pointed out, the adiabatic requirement is not enough to select an unique expression for the modes \eqref{modeansatz}. We impose the condition that the stress-energy tensor vanishes, mode-by-mode, after 
  renormalization \cite{Agullo15}, such that  [see \eqref{rhop}], 
\bea \label{IV-cond}
\langle \rho_k \rangle (\eta_0)=C_\rho(\mu_{*}, k, \eta_0) ,\qquad \langle p _k \rangle(\eta_0) =C_p(\mu_{*}, k, \eta_0).
\eea
Inserting the ansatz \eqref{IV-1} into the mode expansion \eqref{rhok} and \eqref{pk}, we can solve \eqref{IV-cond} for $W_k$ and $V_k$, arriving to
\bea
W_k(\eta_0)&=&\frac{2k^2+3\,Q (\eta_0) a(\eta_0)^2}{6a(\eta_0)^4(C_\rho(\mu_*,k,\eta_0)-C_p(\mu_*,k,\eta_0))} \label{W00}\, ,\\
\nonumber \\
V^{(\pm)}_k(\eta_0)&=&\frac{2a'(\eta_0)}{a(\eta_0)} \pm \sqrt{-W_k(\eta_0)^2-k^2  -a(\eta_0)^2Q(\eta_0)+4 a(\eta_0)^4 C_\rho(\mu_*,k,\eta_0)W_k(\eta_0)} \, . \label{V00}
\eea
It can be argued \cite{Agullo15} that $V^{(+)}_{k}(\eta_0)$ is the appropriated solution in expanding cosmologies while $V^{(-)}_{k}(\eta_0)$ has to be used in the contracting case. Although we have been able to solve \eqref{IV-cond} algebraically, the consistency of the solutions is not ensured. As stressed  above, these solutions will be consistent if and only if $W_k(\eta_0)$ and $V_k(\eta_0)$ are finite and real and $W_k(\eta_0)$ is also positive.  The condition for $V_k(\eta_0)$ translates to 
\bea
\infty >r_k(\eta_0)=-W_k(\eta_0)^2-k^2  -a(\eta_0)^2 Q(\eta_0)+4 a(\eta_0)^4 C_\rho(\mu_*,k,\eta_0)W_k(\eta_0) \geq 0. \label{cond1}
\eea
Both expressions $r_k$ and $W_k$ depend on the expansion parameter $a(\eta)$ and $Q(\eta)$ evaluated at $\eta=\eta_0$ and its four and two first time derivatives respectively, and on $k$ and $\mu_{*}$, so in general it is not trivial to ensure that these conditions are satisfied.\\

\begin{figure}[h!]
    \centering
    \includegraphics[width=0.85\textwidth]{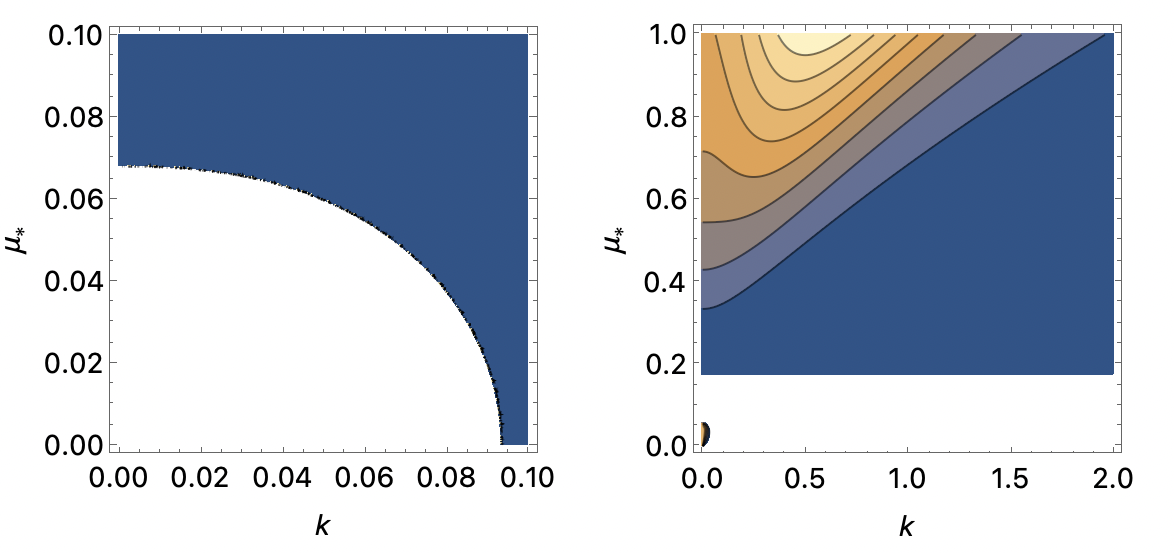} 
    \caption{ We have plotted the value of $W_k$ (left) and $r_k$ (right) for different values of $k$ and $\mu_*$ in the case of $Q=m^2+3\phi_*^2\cos^2{\left[m (\eta-\eta_0)\right]}$ with $m=10^{-4}$, $\phi_*=10^{-1}$ and $a(\eta)=1$ in natural Planck units. The colored patches are the regions where $W_k$ and $r_k$ are positive respectively.}
    \label{fig:mesh2}
\end{figure}

For instance, take the case of a free field, $Q=m^2$. In  \cite{Agullo15} it was proven that for the standard adiabatic regularization prescription ($\mu_*=m$), the conditions are indeed satisfied for 
some physically motivated cosmological models.  However, in the case of a time-dependent effective mass, e.g., $Q=m^2+3\lambda \bar{\phi}^2$, this particular prescription is no longer valid. This can be seen in Fig. \ref{fig:mesh2} where we have represented the magnitudes $W_k$ and $r_k$  for the case of $\lambda\bar{\phi}^2=\phi_*^2\cos^2{\left[m (\eta-\eta_0)\right]}$ with $m=10^{-4}$ $\phi_*=10^{-1}$ in natural Planck units at $\eta=\eta_0$ in the Minkowski limit. There is a minimum value of the parameter $\mu_*=\mu_{\rm min}$ that allows to have both positive $r_k(\eta_0)$ and $W_k(\eta_0)$ for all $k$, but it is a higher value than the standard adiabatic regularization presupposes $\mu_*=m$.\\

A natural question that arises is whether we can find always a $\mu_*$ big enough to make both magnitudes positive irrespective to the possible values of $Q$ and $a$ and its derivatives. The answer is affirmative since the behaviour for large $\mu_*$ is 
\bea \label{conditions-mu-infinity}
W(\eta_0)=\left(\frac{16}{9}k^2+\frac83 a^2(\eta_0)m^2(\eta_0)\right) a^{-1}(\eta_0)\mu_*^{-1}+\mathcal{O}\left(\mu_*^{-3}\right),~~~~~
r(\eta_0)= \frac13 \left(k^2+3a^2(\eta_0)m^2(\eta_0)\right)+\mathcal{O}\left(\mu_*^{-2}\right).
\eea
Note that for $k\to \infty$, the positiveness of both magnitudes is always ensured. This is because, for $k \to \infty$, the modes behave like the adiabatic expansion, as required by the adiabatic or Hadamard condition. Nevertheless, there are two disadvantages of this procedure. First, there is not a unique value for $\mu_*$ such that both conditions hold. Second, even if we choose the minimum value that guarantees these conditions, there is not a simple analytical value of $\mu_{\rm min}$ which makes the potential physical interpretation of the scale complicated. We will show how to compute $\mu_{\rm min}$ in two physically motivated physical models.

\subsection{Example: massless scalar field with a quartic potential during preheating phase}
\begin{figure}[h]
    \centering
    \includegraphics[width=0.85\textwidth]{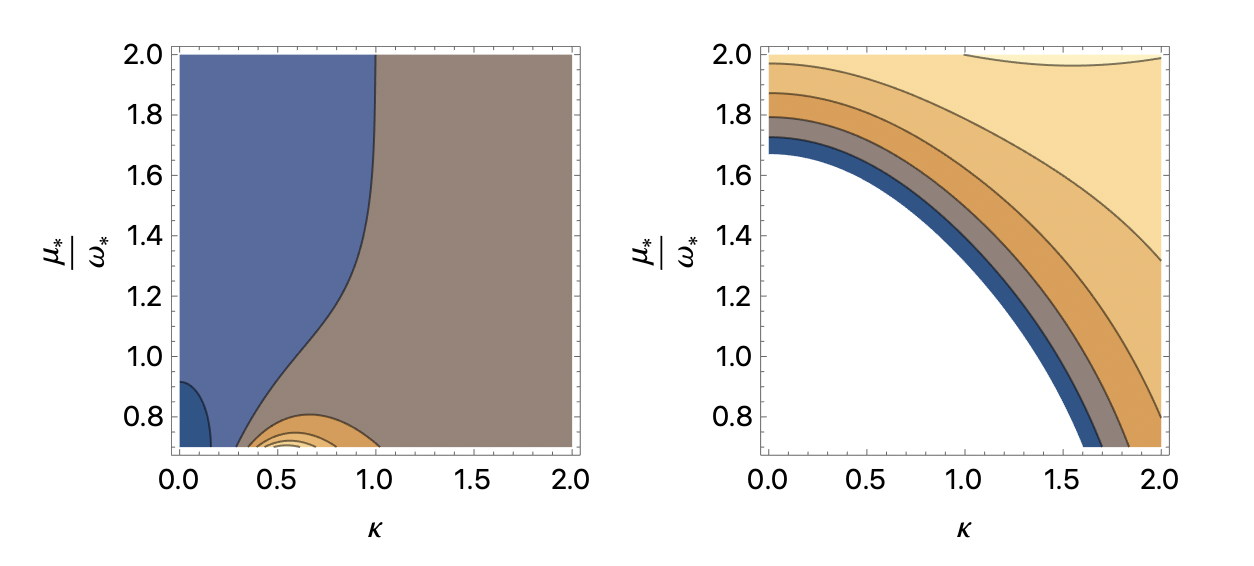}
    \caption{We plot the quantities $\tilde{W}_k$ (left) and $\tilde{r}_k$ (right) for $H_*=0.69\omega_*$ at $z_0=0$. The x-axis is $\kappa$ and the y-axis is $\mu_*$. The colored patches are the regions where $W_k$ and $r_k$ are positive respectively.}
    \label{fig:mesh1-a}
\end{figure}

A typical case commonly used to describe the production of particles during the preheating phase is a massless scalar field, falling down the potential $\lambda \phi^4$ \cite{reheating}, i.e., $Q=3 \lambda \bar \phi^2$. Under some physically motivated approximations, during the first oscillations, the expanding parameter and the classical scalar field take the form (see \cite{turner, paco1,paco2})
\bea
a(z)=a_*\left(1+\frac{H_{*}a_*}{\omega_*}z\right),~~~~~~~\bar \varphi(z)=\text{cd}\left(\frac{z}{\sqrt{2}},-1\right).\label{pacopot}
\eea
Here we have defined a convenient change of variables:
\bea
\eta \to z:= \lambda^{1/2}\phi_{*}\eta\equiv \omega_*\eta, ~~~~~~  \bar{\phi} \to \bar \varphi:=a\phi_{*}^{-1}  \bar{\phi} \label{newvar}
\eea
where $a_*$ and $\phi_{*}$ are the values of the expanding parameter and background scalar field at the initial time $z=0$ respectively. Under this configuration, and in order to construct the instantaneous vacuum, the only free parameters are $H_*$, $\mu_*$, $\phi_*$ and $k$. Furthermore, since we are only interested in the sign of expressions \eqref{W00} and \eqref{cond1} we can re-scale both expressions, $W_k \to  \tilde{W}_k=\omega^{-1}_*W_k $ and $r_k \to  \tilde{r}_k=\omega_*^{-2}r_k$ such that they only depend on $\tilde{H}_*=\omega_*^{-1}H_*$,  $\tilde{\mu}_*=\omega_*^{-1}\mu_*$ and $\kappa=\omega^{-1}_*k$. In Fig. \ref{fig:mesh1-a}, we plot both $\tilde{W}_k$ and $\tilde{r}_k$ for different values of $\kappa$ and $\tilde{\mu}_*$ and for $H_*=0.69 \omega_*$, which is the physical value  obtained by solving the coupled inflaton and scale factor equations numerically  at the end of the inflationary phase \cite{paco1}. We see again, that the condition \eqref{cond1} imposes a minimum value of $\mu_*$, for which the condition of \eqref{W00} is automatically satisfied. \\

\begin{figure}[h]
    \centering
    \includegraphics[width=0.85\textwidth]{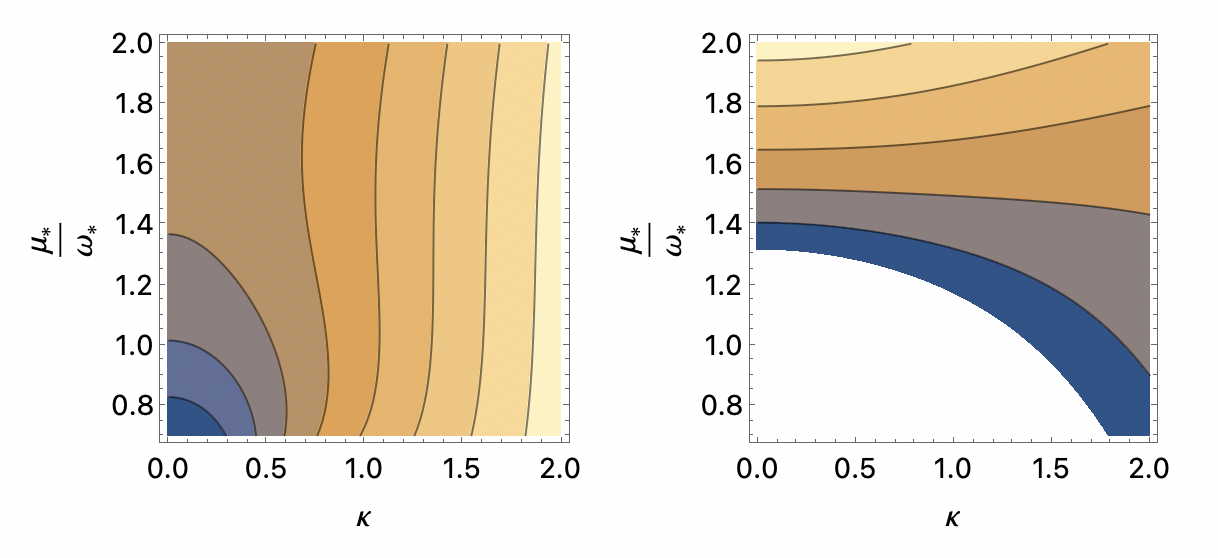}
   \caption{We plot the quantities $\tilde{W}_k$ (left) and $\tilde{r}_k$ (right) for $H_*=0.69\omega_*$ at $z_0=0.5$. The x-axis is $\kappa$ and the y-axis is $\mu_*$. The colored patches are the regions where $W_k$ and $r_k$ are positive respectively.}
    \label{fig:mesh1-b}
\end{figure}

There are several comments in order. First, we have checked that a similar consistent result is obtained for a variety of values for $H_*$, namely that there is always a minimum value for $\mu_*$ that satisfies both the conditions \eqref{W00} and \eqref{cond1}. Second, we have also checked that this result is independent of the vanishing value of $\bar \varphi'$. 
This can be seen in Fig. \ref{fig:mesh1-b} where we have plotted the same expressions as in Fig. \ref{fig:mesh1-a} but for $z=0.5$ where both $\bar \varphi$ and $\bar \varphi'$ have a non-vanishing value. This result confirms the robustness of this method. \\

\begin{figure}[h]
    \centering
    \includegraphics[width=0.8\textwidth]{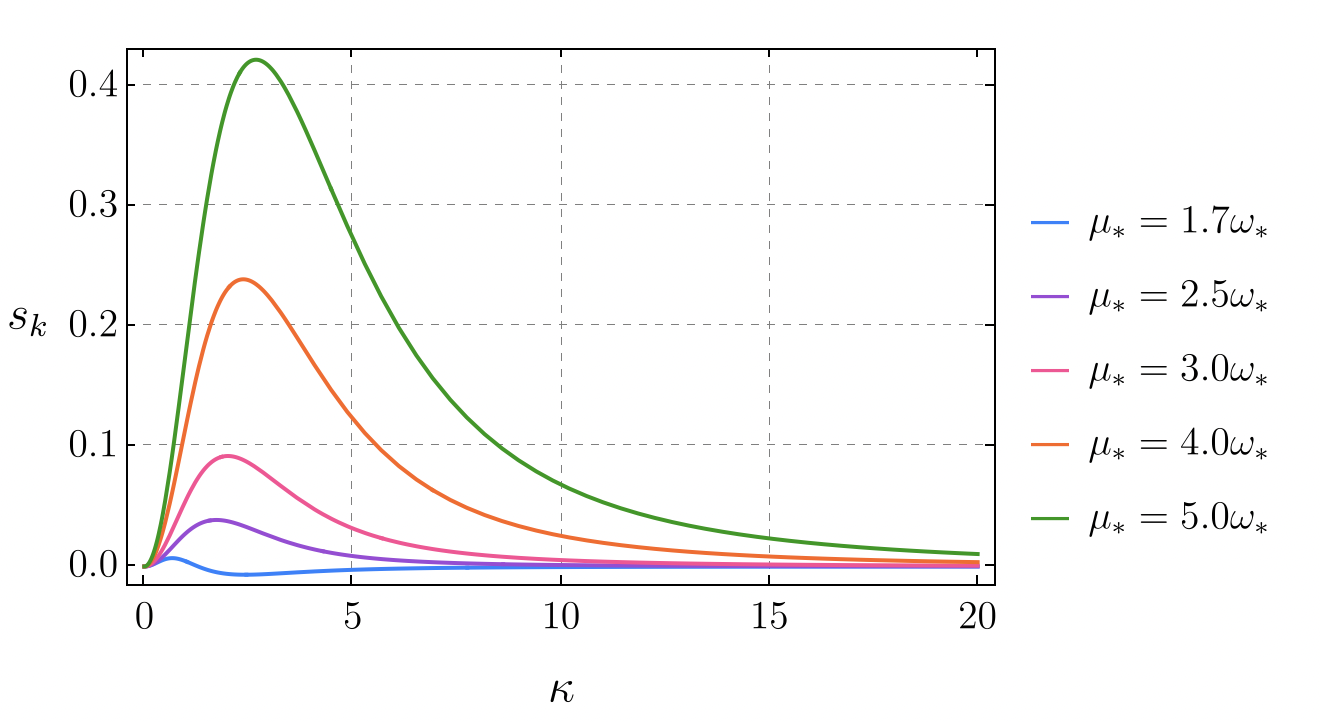}
    \caption{Value of $s_{\kappa}$ defined in \eqref{tpfsk} for different values of $\mu_*$ and $\tilde H_*=0.69$.  }
    \label{fig:sk}
\end{figure}

Finally, it is tempting to associate the minimum value for $\mu_*$ in Fig. \ref{fig:mesh1-a} with some combination of physical values. Let us take a closer look into this particular case. The equation of motion of $h_{k}$ in terms of the new variables \eqref{newvar} takes the form 
\bea
\frac{\partial^2}{\partial z^2}\left(a(z) h_{k}(z)\right)+\left(\kappa^2+3 \bar \varphi^2(z)+(6\xi-1)\frac{a(z)''}{a(z)}\right)\left(a(z) h_{k}(z)\right)=0.
\eea
For our specific case \eqref{pacopot}, the term involving the time derivative of the expanding parameter vanishes for all $z$. The only physical relevant scale would be then $\bar \varphi(z)$. In this case the most clear assignation would be $\tilde{\mu}_* \sim \bar \varphi(z)$. However, since we have constructed the vacuum state from the modes from the stress-energy tensor, additional time derivatives appear, and therefore both $W_k$ and $r_k$ will also depend on $H_*$. A possible combination of both magnitudes to generate this minimum value would be rather unnatural. Since it is not possible to find a trivial expression for the minimum value of $\mu_{\rm min}$, there is still freedom of choosing any other $\mu_*>\mu_{\rm min}$. In order to decide which value is the optimal, it is useful to construct the remainder magnitude appearing in the semiclassical equations, i.e., the renormalized two-point function at $\eta_0$
\bea
\langle \delta \phi^2\rangle_{\rm ren}(\eta_0)=\frac{1}{2\pi^2}\int dk k^2 \frac{1}{2a^2}W^{-1}_k(\eta_0)-\Phi^{(0)}(\eta_0)-\Phi^{(2)}(\eta_0)\equiv \frac{ \omega_*^2}{4a^2\pi^2}\int s_{\kappa} d\kappa,\label{tpfsk}
\eea
where $\Phi^{(0)}$ and $\Phi^{(2)}$ are defined in \eqref{phiadb}. Here $s_{\kappa}$ only depends on $\tilde H_*$, $\kappa$ and $\tilde \mu_*$. In Fig. \ref{fig:sk} we plot $s_{\kappa}$ for several permitted values of $\tilde \mu_*$ and $\tilde H_*=0.69$. We see that if we increase the value of $\mu_*$, the fluctuations at this initial time also increase.
The initial value of the two point function is very important since it enters explicitly into the semiclassical equations for the classical scalar field
\bea
\bar \varphi''+\bar \varphi^3+(3\lambda a^2 \langle \delta \phi^2 \rangle_{\rm ren})\bar \varphi=0 .
\eea
In the next section, we will see how the minimum value of $\mu_*$ can prevent the backreaction to be very strong at the initial time, a desirable feature when computing the quantum fluctuations and the backreaction during preheating.

\section{Instantaneous vacuum for a daughter field with a quadratic potential} \label{sec:daughterfield}
 Another widely used model in the context of preheating consists of a massless quantum scalar field   $X$ (the daughter field) coupled to the classical massive scalar inflaton field via $\frac12 g^2\phi^2 X^2$. It is easy to check that, the quantization is equivalent to the one carried out in Section \ref{sec:lambdaphi4}. 
 The quantized field $X$ can be expanded in terms of modes $h_k$ as in \eqref{deltaphimod}, but now obeys equation \eqref{eom} with $Q=g^2 \bar \phi^2 (\eta)$ (For a more detailed description see for example \cite{Anderson1}). 
We can also define the two-point function, the energy density and the pressure as in Eqs. \eqref{eq:tpf} and \eqref{Trhop} respectively. In this context, it is convenient to introduce the following natural variables  
\be \eta \rightarrow z := \int_{\eta_{0}}^\eta {\omega_{*} \, a(\eta')d\eta'} \hspace{0.4cm} \varphi := a^{3/2} \phi_{*}^{-1}   \phi \ , \hspace{0.4cm} \chi_k :=  a^{3/2} \phi_{*}^{-1}  h_{k} \ , \label{eq:newvars1}\ee
with $\omega_{*} \equiv   m $. With these changes, the equation for the daughter field modes $\chi_k$ becomes (assuming $\xi=0$)  
\be
\chi''_k + (\Delta+\frac{\kappa^2}{a^2}+ Q_{*})\chi_k=0\, ,
\ee
where $\kappa=\omega_{*}^{-1}k$, $\Delta(z)=-\frac{3}{4}\big(\frac{a'}{a}\big)^2-\frac{3}{2}\frac{\,a''}{a}$ and $Q_{*}=\frac{g^2 \phi_{*}^2}{\omega_{*}^2a^3}\bar \varphi^2\equiv q\, a^{-3}\bar \varphi^2 $ together with the Wronskian condition $\chi_k \chi^{*\prime}_{k} -\chi^*_k \chi'_k=i \phi_{*}^{-2} \omega_*^{-1}$.\footnote{Here, the prime refers to the derivative with respect to $z$.} The equation for the background, ignoring backreaction effects, reads 
\bea
\bar \varphi''+(\Delta(z) +1)\bar \varphi=0.
\eea

Ignoring the backreaction of the daughter field at this point (we will come to this issue later on), we can solve the semiclassical equations during the first oscillations around the potential and obtain \cite{paco1,paco2}
\be
a(z)=\Big(1+ \frac{3}{2}\frac{H_{*}}{\omega_{*}} z\Big)^{2/3},~~~~~
\bar \varphi=\cos (z)\, ,\label{reheatingdaughter}
\ee
which leads to $\Delta=0$. Here we have fixed the initial conditions $\bar \varphi(0)=1$ and  $\bar \varphi'(0)=0$.\\

Again, the vev of the stress-energy tensor and the two-point function of the quantized field $X$ diverge, and adiabatic regularization can be used to obtain finite results. We will not repeat the process again since it is straightforward using the method explained in Section \ref{sec:adiabatic}, by fixing $Q=g^2\bar \phi^2$, $m=0$ and $\xi=0$. The subtraction terms can be found in Appendix \ref{app:subtractions}. \\

In this context, the instantaneous vacuum for the daughter field can be defined exactly as before \eqref{IV-1}, and conveniently re-expressed in terms of the new variables introduced in \eqref{eq:newvars1}. We can also scale the frequency  as $\omega \to \tilde\omega = a^{-1}\omega^{-1}_{*}\omega\equiv\sqrt{\kappa^2a^{-2} + \tilde \mu_{*}^2}$ where $\tilde \mu_{*}= \omega^{-1}_{*}\mu_{*}$. Therefore, the re-scaled functions $\tilde W_k = a^{-1} \omega_{*}^{-1}W_k$, $\tilde V_k= a^{-1} \omega_{*}^{-1}V_k$ ,  and $\tilde r_k=a^{-2} \omega_{*}^{-2}r_k$ depend only on $q$, $\kappa$, 
$\tilde \mu_{*}$ and $\tilde H_{*}=\omega_{*}^{-1}H_{*}$, 
\bea
\tilde W_k(z_0)&=&\frac{2 a(z_0)^{-2}\kappa^2+3\,Q_* (z_0)}{6a(z_0)^3 \omega_*^{-1}(C_\rho(\tilde \mu_*,\kappa,z_0)-C_p(\tilde \mu_*,\kappa,z_0))} \label{W00-bis}\, ,\\
\nonumber \\
\tilde V^{(\pm)}_k(z_0)&=&\frac{2a'(z_0)}{a(z_0)} \pm \sqrt{-\tilde W_k(\eta_0)^2-a(z_0)^{-2}\kappa^2 -Q_*(z_0)+4 a(z_0)^3 \omega_*^{-1} C_\rho(\tilde \mu_*,\kappa,z_0)\tilde W_k(z_0)} \, . \label{V00-bis}
\eea

\begin{figure}[h!]
    \centering
      \centering
    \includegraphics[width=0.85\textwidth]{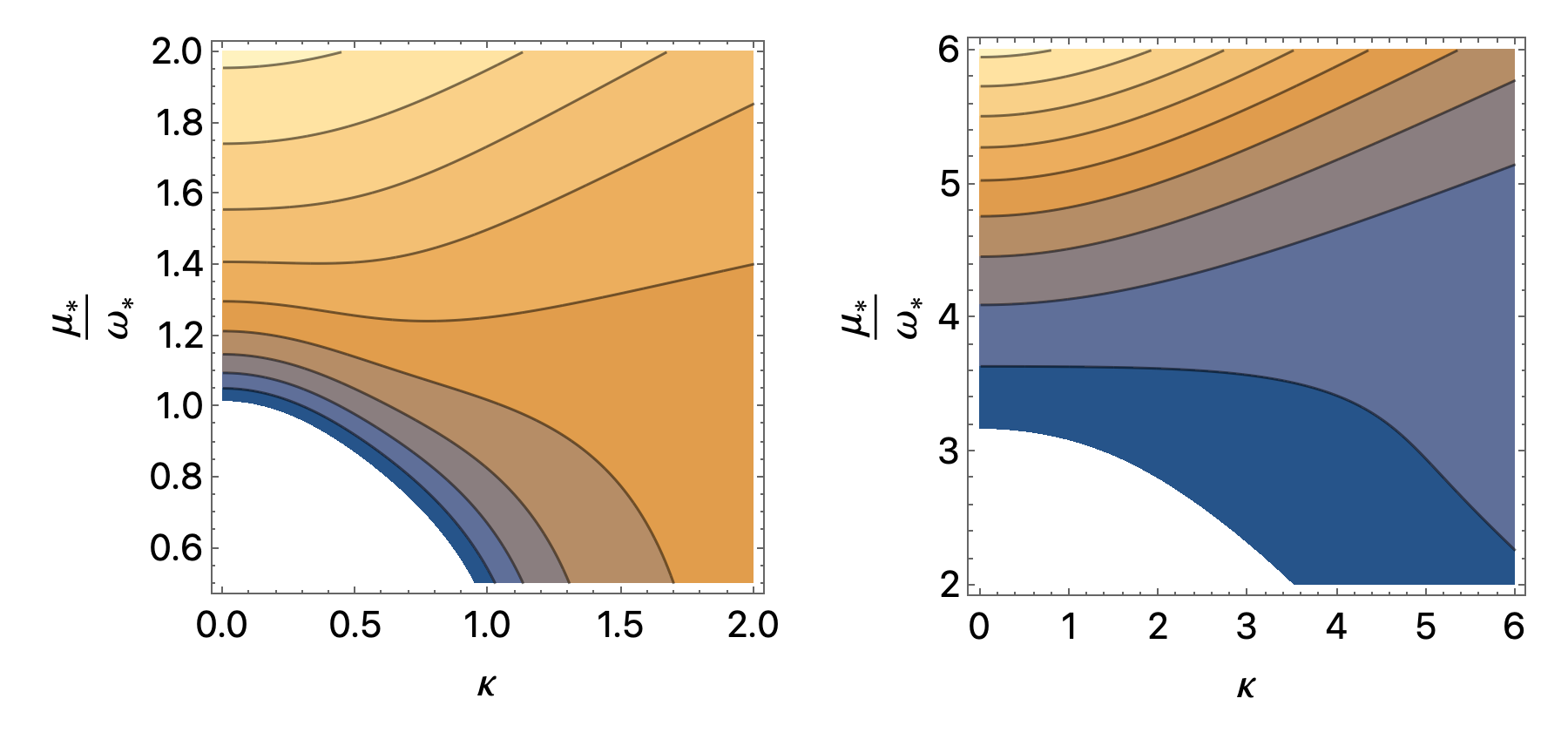}\\
      \centering
    \includegraphics[width=0.85\textwidth]{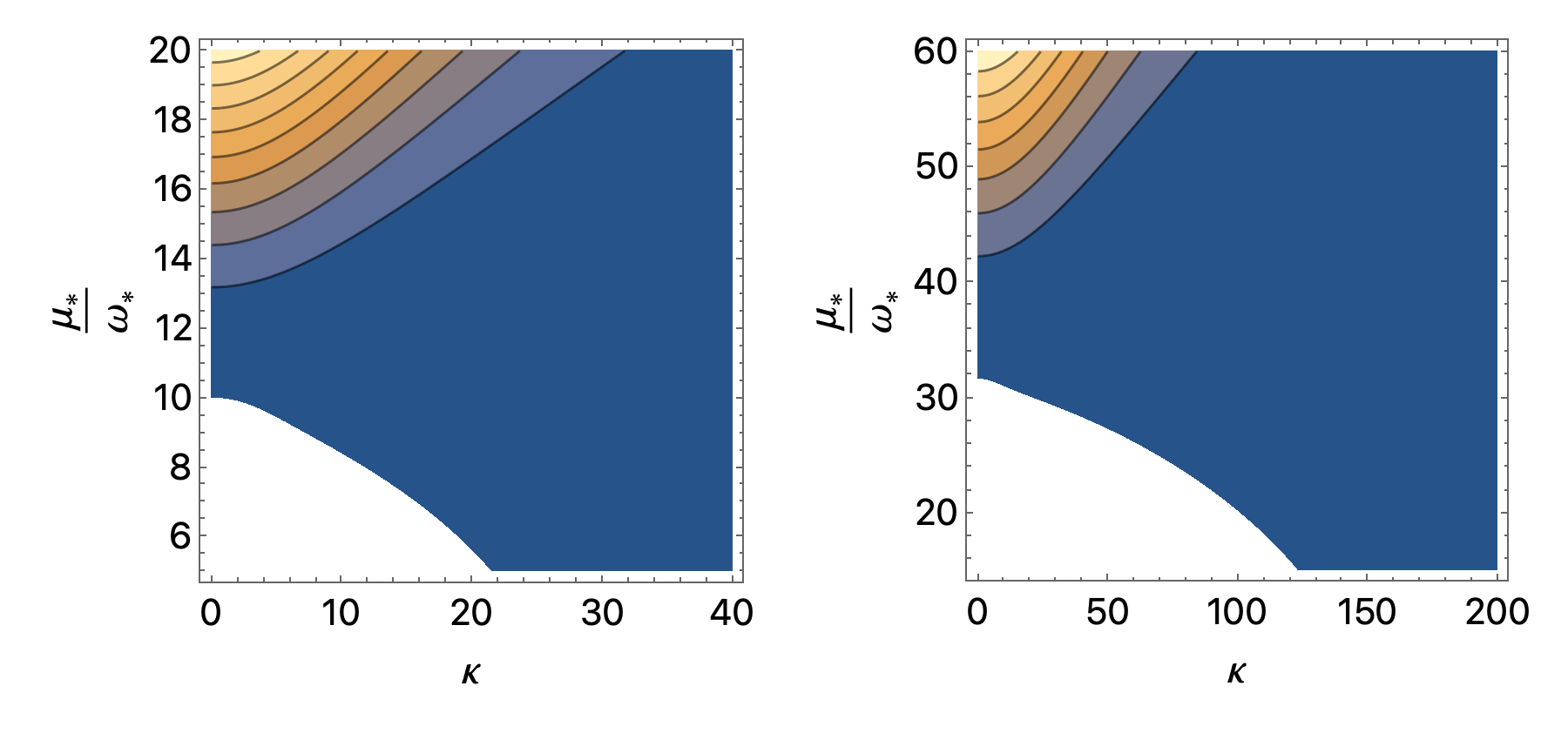}
    \caption{\small{We plot the quantity $\tilde{r}_k$ for $q=1$, $q=10$ (first row), $q=10^3$ and $q=10^4$ (second row). The value of $\tilde H_{*}$ is fixed to be $\tilde H_{*}=0.5$. The colored patches are the regions where $W_k$ and $r_k$ are positive respectively.}}
    \label{fig:mesh-daughter}
\end{figure}

We aim to explore the values of $\tilde \mu_{*}$ for which the instantaneous vacuum is well-defined, i.e., $\tilde W_{k}>0$ and $\tilde r_{k}\geq0$. For this example, we will only focus on $\tilde r_k$ since it can be checked that for all the studied cases, the condition on  $\tilde r_k$ is more restrictive than the one imposed on $\tilde W_{k}$. The value of  $\tilde H_{*}$ can be computed by solving Einstein's equations during the first oscillations of the scalar field and yields  $\tilde H_{*}=0.5$ \cite{paco1}. In Fig. \ref{fig:mesh-daughter}  we represent $\tilde r_k$ for different values of $\kappa$ and $\tilde \mu_{*}$. We have repeated the analysis for different values of $q$. \\

As in the $\lambda \phi^4$ model, not all values of $\mu_*$ are allowed for constructing the modes of the instantaneous vacuum consistently.In Fig. \ref{fig:mesh-daughter} we also see that the minimum value required is $\tilde \mu_{\textrm{min}}\sim \sqrt{q}$. However, we note that in this example, the choice $\tilde \mu_{*}=\sqrt q$ does not satisfy the requirement $\tilde r_{k}>0$. That is, $\tilde \mu_{\textrm{min}}$ should be greater (although sometimes it is very close) than $\sqrt q$. In general, it will be a non-trivial combination of the parameters $q$ and $\tilde H_{*}$. 
However, as we noted in the case of $\lambda \phi^4$, increasing the value of $\mu_*$ increases the value of $\langle X^2 \rangle_{\rm ren}$ for the instantaneous vacuum state at $\eta=\eta_0$. As we will see, this can have dramatic consequences in the backreaction to the classical scalar field, and the choice of a consistent value of $\mu_*$ has to be done carefully.

\subsubsection{Backreaction effects}

For simplicity let us restrict to Minkowski spacetime. The equation of motion for the background field $\bar \varphi$ including backreaction effects is
\be \label{eq:backreaction-daughter}
\bar \varphi'' + (1+g^2 \langle \text{x}^2\rangle_\textrm{ren})\bar \varphi=0\, ,
\ee
where $\langle \text{x}^2\rangle_\textrm{ren}$ is the renormalized two-point function for the daughter field $\langle X^2\rangle_{\textrm{ren}}$ [see Eq. \eqref{eq:tpf}] in terms of the re-scaled variables, which at $z=0$ for the instantaneous vacuum \eqref{IV-1} becomes 
\be
\langle \text{x}^2_0 \rangle_{\textrm{ren}}=\Big(3(\tilde\mu_{*}^2-q\bar \varphi_0^2)^2+2 q\bar \varphi_0^{\prime2}+2 q \bar \varphi_0\bar \varphi''_0\Big)\int \frac{d^3\kappa}{(2 \pi)^3} S_{\kappa}  \equiv \frac{1}{2 \pi^2}\, \Big(3(\tilde\mu_{*}^2-q\bar \varphi_0^2)^2+2 q\bar \varphi_0^{\prime2}+2 q \bar \varphi_0\bar \varphi''_0\Big) 
I(\tilde \mu_*,q,\bar \varphi) \label{x2} \,\ee
where $S_\kappa=\frac{(2\kappa^2+3 \tilde \mu_*^2)}{16 \tilde \omega^5 (2\kappa^2+3 q\bar \varphi_0^2)}$, and $ \bar \varphi_0$, $\bar  \varphi'_0$ and $\bar \varphi''_0$ refer to the initial values of the background $\bar \varphi(z)$ and its two first derivatives respectively. From the expression above and using \eqref{eq:backreaction-daughter} we find 
\be \label{eq:ddphi-0}
\bar \varphi''_0=-\frac{1+\frac{g^2}{2\pi^2}\,I\,\big[3(\tilde \mu_{*}^2-q\bar \varphi_0^2)^2+2 q \bar \varphi_0^{\prime 2}\big]}{1+\frac{g^2}{\pi^2}\,I\,  q \bar\varphi_0^2} \bar \varphi_0.
\ee

In order to quantify the backreaction of the produced quantum fluctuations during a finite amount of time, e.g., during the first oscillations of the background scalar field of the preheating phase, it is necessary to ensure that the quantum fluctuations are not very large at the initial time, such that the classical solutions \eqref{reheatingdaughter} are valid. This is, in general, not an easy problem since, for possible physical values of the couplings and the scalar field, the backreaction effects can become very large at initial times. For example, in \cite{Anderson1} a possible vacuum state was proposed and the two-point function $\langle \text{x}^2\rangle_\textrm{ren}$ was calculated. It is not difficult to check that this vacuum state can be recovered from the instantaneous vacuum by choosing $\mu_*^2=q \varphi_0^2$. In this case, it was proven that possible values of $q$ can produce a large value of quantum fluctuations, and therefore only certain values are optimal to study the backreaction effect. \\

In the case of the instantaneous vacuum, this issue can be overcome, and we can construct a vacuum state that has small fluctuations, given any possible values of $g$ and $q$. To see this, we assume the initial conditions $\bar \varphi_0=1$ and $\bar \varphi'_0=0$. In this case, the initial backreaction contribution can be completely suppressed (i.e., $\varphi''_0=-1$) by fixing
\be
\tilde \mu^2_*=q+\sqrt{\tfrac{2}{3}}\sqrt{q}\, .\label{qvalue}
\ee
On the other hand, it can be easily shown that as long as $\tilde \mu_*$ starts to increase, the backreaction effects grow without bound for arbitrarily large values of $\tilde \mu_*$. In fact, the asymptotic behavior of $\varphi_0''$ for large  $\tilde \mu_*$ is \eqref{eq:ddphi-0}
\be
\varphi_0'' =-\frac{g^2}{8 \pi^2}\, \tilde \mu_*^2+\mathcal O(\tilde \mu_*)\, .
\ee
We conclude that $\tilde \mu_*$ has to be large enough to ensure $\tilde W_k >0$ and $\tilde r_k \geq 0$, but small enough to minimize the initial backreaction effects.



\section{Conclusions and further comments}
\label{sec:conclusions}
In this work, we have generalized the extended adiabatic regularization to an interacting $\lambda \phi^4$ theory at one-loop order. This extended version includes an arbitrary mass parameter $\mu$. We have used this arbitrariness to consistently construct a natural instantaneous vacuum state, requiring that the stress-energy tensor vanish mode-by-mode. An important result is that not all values of $\mu$ are admissible to construct this vacuum state.\\

We have constructed the vacuum state for two well-known models in cosmology, in particular during the preheating stage after the inflationary period, namely: the massless $\lambda \phi^4$ and the daughter field coupled to a massive scalar field with quadratic potential. A significant result is that we can tune the value of $\mu$ such that the initial fluctuations encoded in $\langle\delta \phi^2 \rangle_{\rm ren}$ are not too strong, a feature highly desirable for computing the backreaction of the produced quanta during the preheating phase. \\

Finally, we would like to comment on another interesting advantage of the instantaneous vacuum. Assume we have constructed a solution for the quantized scalar field, e.g., the daughter field described in Section \ref{sec:daughterfield}, in terms of the modes $h_k(\eta)$ imposing the instantaneous vacuum condition \eqref{IV-cond} at some initial time $\eta_0$. We can always construct another mode sum expansion for the scalar field in terms of $\tilde{h}_k(\eta)$ which 
satisfies  \eqref{IV-cond} at a different time $\eta_f$. Both solutions are related in the following way:
\bea
h_k(\eta)=\alpha_k \tilde{h}_k(\eta)+\beta_k \tilde{h}^*_k(\eta)\, ,\\
h'_k(\eta)=\alpha_k \tilde{h}'_k(\eta)+\beta_k \tilde{h}^{*\prime}_k(\eta)\, ,
\eea
where $\alpha_k$ and $\beta_k$ are time-independent coefficients, normalized as $|\alpha_k|^2-|\beta_k|^2=1$ (see, for example, \cite{Ford21}). Note that the existence of $\tilde{h}_k$ will depend on the value $\mu_*$ used to construct the subtraction terms. One can check that for the typical examples of a background scalar field, the value of $\mu_*$ used to construct $h_k$ also serves to construct $\tilde{h}_k$. For simplicity, we restrict to Minkowski spacetime, but the same argument can be made for non-flat spacetimes. The renormalized energy density of the modes $h_k$ at 
$\eta_f>\eta_0$ can be written in terms of $\tilde{h}_k$ as
\bea
\langle \rho \rangle_{\rm ren}(\eta_f)=\frac{1}{(2\pi)^3}\int d^3k\left(2|\beta_k|^2C_{\rho}(\mu_*,k,\eta_f)+2Re(\alpha_k\beta^*_k(\tilde{h}'_k(\eta_f)^2+(k^2+g^2\bar{\phi}^2)\tilde{h}_k(\eta_f)^2)\right),\label{rhon2}
\eea
with 
\bea
\beta_k=\frac{i}{2}\left(\tilde{h}'_k(\eta_f)h_k(\eta_f)-\tilde{h}_k(\eta_f)h_k'(\eta_f)\right).
\eea
A similar expression holds for the pressure. The advantage of this formulation is that, by construction, \eqref{rhon2} is a finite quantity. We can interpret here the $|\beta_k|^2$ as a measure of the number of particles carrying an energy density $C_{\rho}$, which is nothing 
but the energy density subtracted from the exact energy density at initial time $\eta_0$. However, we must be careful with this interpretation since another definition for the vacuum state, i.e., another construction of $\tilde{h}_k$ would yield another result for $|\beta_k|^2$, meaning that the interpretation is not unique. Only in particular cases where a unique natural definition of the vacuum state is available this can be done. For example, consider that after some time $\eta_f$, the classical field $\bar{\phi}\approx\bar{\phi}_f$ becomes constant. In this case, there is a natural vacuum state of the form 
\bea
f_k(\eta)=\frac{1}{\sqrt{2\Omega_k}}~~~f'_k(\eta)=- i \sqrt{\frac{\Omega_k}{2}}\label{vacuummink}
\eea
where here $\Omega_k=\sqrt{k^2+g^2\bar{\phi}_f^2}$. 
This vacuum can be recovered from the instantaneous vacuum by choosing the mass scale to be $\mu_f^2=g^2\bar{\phi}_f^2$.  
However, we have already fixed the $\mu_*$ at the initial time $\eta_0$, e.g. using \eqref{qvalue}. We can reformulate 
the energy density in terms of the vacuum \eqref{vacuummink} as
\bea
\langle \rho \rangle_{\rm ren}(\eta_f)=\frac{1}{(4\pi^2)}\int_0^{\infty} dkk^2|\beta_k|^2\Omega_k+\rho_{\rm vac}(\mu_*,\bar{\phi}_f)\label{rhon4}
\eea
where  
\bea
\rho_{\rm vac}(\mu_*,\bar{\phi}_f)=\frac{1}{128\pi^2}\left(-\mu_*^4+4\mu_*^2g^2\bar{\phi}_f^2-3g^4\bar{\phi}^4_f-\frac12\log{\left(\frac{g^2\bar{\phi}_f^2}{\mu_*^2}\right)}\bar{\phi}_f^4\right),
\eea
We see a clear distinction between the contribution to the energy density associated with the particle production and the energy density associated with the vacuum $\rho_{\rm vac}$. Note that, even if the particle production were negligible $|\beta_k|^2\approx 0$, since the energy density at the initial time $\eta_0$ is zero, we would have a net production of vacuum energy due to the change of the classical scalar field. Of course, the possibility of choosing a unique vacuum state is not guaranteed in a non-flat spacetime or even when the classical scalar field is non-constant. However, the formulation proposed in this work allows us to quantify both particle production and vacuum polarization effects in a transparent way, using, for example, \eqref{rhon2}. Applying this formulation to compute the backreaction of preheating processes is the motivation for future projects. 

\section*{Acknowledgements}
The authors thank J. Navarro-Salas and F. Torrenti for his support and useful discussions.  Part of this work is supported by the Spanish Grant 
 PID2020-116567GB-C2-1 funded by MCIN/AEI/10.13039/501100011033. 
A. F. is supported by the Irish Research Council Postdoctoral Fellowship No. GOIPD/2021/544. S. P. is supported by the Ministerio de Ciencia, Innovaci\'on y Universidades, Ph.D. fellowship, Grant No. FPU16/05287.

\appendix

\section{Adiabatic subtractions of the stress-energy tensor $C_p$ and $C_\rho$}  \label{app:subtractions}
Defining $s=Q-\mu^2$ and using $\omega=\sqrt{k^2+a^2 \mu^2}$, the adiabatic subtractions for the energy density and pressure up to and including the fourth adiabatic order read
\bea
C_\rho=&&\frac{\omega }{2 a^4}+\frac{s}{4 a^2 \omega }+\frac{\left(a'\right)^2}{4
   a^6 \omega }-\frac{3 \xi  \left(a'\right)^2}{2 a^6 \omega }+\frac{a'
   \omega '}{4 a^5 \omega ^2}-\frac{3 \xi  a' \omega '}{2 a^5 \omega
   ^2}+\frac{\left(\omega '\right)^2}{16 a^4 \omega ^3}-\frac{s \sigma }{8 a^2 \omega ^3}+\frac{\sigma ^2}{16 a^4 \omega
   ^3}\\
   &&-\frac{\sigma  \left(a'\right)^2}{8 a^6 \omega ^3}+\frac{3 \xi 
   \sigma  \left(a'\right)^2}{4 a^6 \omega ^3}+\frac{a' \sigma '}{8 a^5
   \omega ^3}-\frac{3 \xi  a' \sigma '}{4 a^5 \omega ^3}
   -\frac{3 \sigma 
   a' \omega '}{8 a^5 \omega ^4}+\frac{9 \xi  \sigma  a' \omega '}{4 a^5
   \omega ^4}+\frac{\sigma ' \omega '}{16 a^4 \omega ^4}-\frac{3 s
   \left(\omega '\right)^2}{32 a^2 \omega ^5}\nonumber\\
   &&-\frac{\sigma  \left(\omega
   '\right)^2}{16 a^4 \omega ^5}-\frac{3 \left(a'\right)^2 \left(\omega
   '\right)^2}{32 a^6 \omega ^5}+\frac{9 \xi  \left(a'\right)^2
   \left(\omega '\right)^2}{16 a^6 \omega ^5}-\frac{15 a' \left(\omega
   '\right)^3}{32 a^5 \omega ^6}+\frac{45 \xi  a' \left(\omega
   '\right)^3}{16 a^5 \omega ^6}-\frac{45 \left(\omega '\right)^4}{256 a^4
   \omega ^7}\nonumber\\
   &&+\frac{s \omega ''}{16 a^2 \omega ^4}-\frac{\sigma  \omega
   ''}{16 a^4 \omega ^4}+\frac{\left(a'\right)^2 \omega ''}{16 a^6 \omega
   ^4}-\frac{3 \xi  \left(a'\right)^2 \omega ''}{8 a^6 \omega ^4}+\frac{7
   a' \omega ' \omega ''}{16 a^5 \omega ^5}-\frac{21 \xi  a' \omega '
   \omega ''}{8 a^5 \omega ^5}+\frac{5 \left(\omega '\right)^2 \omega
   ''}{32 a^4 \omega ^6}\nonumber\\
   &&+\frac{\left(\omega ''\right)^2}{64 a^4 \omega
   ^5}-\frac{a' \omega ^{(3)}}{16 a^5 \omega ^4}+\frac{3 \xi  a' \omega
   ^{(3)}}{8 a^5 \omega ^4}-\frac{\omega ' \omega ^{(3)}}{32 a^4 \omega
   ^5}\, .\nonumber
\eea
\bea
C_p= &&\frac{k^2}{6 a^4 \omega }+\frac{\mu ^2 \sigma }{12 a^2 \omega
   ^3}-\frac{s}{4 a^2 \omega }+\frac{s \xi }{a^2 \omega }+\frac{\sigma }{6
   a^4 \omega }-\frac{\xi  \sigma }{a^4 \omega
   }+\frac{\left(a'\right)^2}{4 a^6 \omega }-\frac{3 \xi 
   \left(a'\right)^2}{2 a^6 \omega }+\frac{a' \omega '}{4 a^5 \omega
   ^2}\\
   &&-\frac{3 \xi  a' \omega '}{2 a^5 \omega ^2}+\frac{\mu ^2
   \left(\omega '\right)^2}{16 a^2 \omega ^5}+\frac{3 \left(\omega
   '\right)^2}{16 a^4 \omega ^3}-\frac{\xi  \left(\omega '\right)^2}{a^4
   \omega ^3}-\frac{\xi  a''}{a^5 \omega }+\frac{6 \xi ^2 a''}{a^5 \omega
   }-\frac{\mu ^2 \omega ''}{24 a^2 \omega ^4}-\frac{\omega ''}{12 a^4
   \omega ^2}\nonumber\\
   &&+\frac{\xi  \omega ''}{2 a^4 \omega ^2}-\frac{\mu ^2 \sigma ^2}{16 a^2 \omega ^5}+\frac{s \sigma }{8 a^2 \omega^3}-\frac{s \xi  \sigma }{2 a^2 \omega ^3}-\frac{\sigma ^2}{16 a^4
   \omega ^3}+\frac{\xi  \sigma ^2}{2 a^4 \omega ^3}-\frac{\sigma 
   \left(a'\right)^2}{8 a^6 \omega ^3}+\frac{3 \xi  \sigma 
   \left(a'\right)^2}{4 a^6 \omega ^3}\nonumber\\
   &&+\frac{a' \sigma '}{8 a^5 \omega
   ^3}-\frac{3 \xi  a' \sigma '}{4 a^5 \omega ^3}-\frac{3 \sigma  a'
   \omega '}{8 a^5 \omega ^4}+\frac{9 \xi  \sigma  a' \omega '}{4 a^5
   \omega ^4}+\frac{5 \mu ^2 \sigma ' \omega '}{48 a^2 \omega ^6}+\frac{13
   \sigma ' \omega '}{48 a^4 \omega ^4}-\frac{3 \xi  \sigma ' \omega '}{2
   a^4 \omega ^4}+\frac{3 s \left(\omega '\right)^2}{32 a^2 \omega ^5}\nonumber\\
   &&-\frac{25 \mu ^2 \sigma  \left(\omega '\right)^2}{96 a^2
   \omega ^7}-\frac{3
   s \xi  \left(\omega '\right)^2}{8 a^2 \omega ^5}-\frac{7 \sigma 
   \left(\omega '\right)^2}{12 a^4 \omega ^5}+\frac{27 \xi  \sigma 
   \left(\omega '\right)^2}{8 a^4 \omega ^5}-\frac{3 \left(a'\right)^2
   \left(\omega '\right)^2}{32 a^6 \omega ^5}+\frac{9 \xi 
   \left(a'\right)^2 \left(\omega '\right)^2}{16 a^6 \omega ^5}\nonumber\\
   &&-\frac{15
   a' \left(\omega '\right)^3}{32 a^5 \omega ^6}+\frac{45 \xi  a'
   \left(\omega '\right)^3}{16 a^5 \omega ^6}-\frac{105 \mu ^2
   \left(\omega '\right)^4}{256 a^2 \omega ^9}-\frac{255 \left(\omega
   '\right)^4}{256 a^4 \omega ^7}+\frac{45 \xi  \left(\omega '\right)^4}{8
   a^4 \omega ^7}+\frac{\xi  \sigma  a''}{2 a^5 \omega ^3}\nonumber\\
   &&-\frac{3 \xi ^2
   \sigma  a''}{a^5 \omega ^3}+\frac{3 \xi  \left(\omega '\right)^2 a''}{8
   a^5 \omega ^5}-\frac{9 \xi ^2 \left(\omega '\right)^2 a''}{4 a^5 \omega
   ^5}-\frac{\mu ^2 \sigma ''}{48 a^2 \omega ^5}-\frac{\sigma ''}{24 a^4
   \omega ^3}+\frac{\xi  \sigma ''}{4 a^4 \omega ^3}+\frac{5 \mu ^2 \sigma
    \omega ''}{48 a^2 \omega ^6}\nonumber\\
   &&-\frac{s \omega ''}{16 a^2 \omega
   ^4}+\frac{s \xi  \omega ''}{4 a^2 \omega ^4}+\frac{7 \sigma  \omega
   ''}{48 a^4 \omega ^4}-\frac{\xi  \sigma  \omega ''}{a^4 \omega
   ^4}+\frac{\left(a'\right)^2 \omega ''}{16 a^6 \omega ^4}-\frac{3 \xi 
   \left(a'\right)^2 \omega ''}{8 a^6 \omega ^4}+\frac{7 a' \omega '
   \omega ''}{16 a^5 \omega ^5}\nonumber\\
   &&-\frac{21 \xi  a' \omega ' \omega ''}{8 a^5
   \omega ^5}+\frac{35 \mu ^2 \left(\omega '\right)^2 \omega ''}{64 a^2
   \omega ^8}+\frac{5 \left(\omega '\right)^2 \omega ''}{4 a^4 \omega
   ^6}-\frac{115 \xi  \left(\omega '\right)^2 \omega ''}{16 a^4 \omega
   ^6}-\frac{\xi  a'' \omega ''}{4 a^5 \omega ^4}+\frac{3 \xi ^2 a''
   \omega ''}{2 a^5 \omega ^4}\nonumber\\
   &&-\frac{5 \mu ^2 \left(\omega ''\right)^2}{64
   a^2 \omega ^7}-\frac{9 \left(\omega ''\right)^2}{64 a^4 \omega
   ^5}+\frac{7 \xi  \left(\omega ''\right)^2}{8 a^4 \omega ^5}-\frac{a'
   \omega ^{(3)}}{16 a^5 \omega ^4}+\frac{3 \xi  a' \omega ^{(3)}}{8 a^5
   \omega ^4}-\frac{5 \mu ^2 \omega ' \omega ^{(3)}}{48 a^2 \omega
   ^7}\nonumber\\
   &&-\frac{23 \omega ' \omega ^{(3)}}{96 a^4 \omega ^5}+\frac{11 \xi 
   \omega ' \omega ^{(3)}}{8 a^4 \omega ^5}+\frac{\mu ^2 \omega ^{(4)}}{96
   a^2 \omega ^6}+\frac{\omega ^{(4)}}{48 a^4 \omega ^4}-\frac{\xi  \omega
   ^{(4)}}{8 a^4 \omega ^4}\, . \nonumber
\eea

\end{document}